# Governing online goods: Maturity and formalization in Minecraft, Reddit, and World of Warcraft communities

SETH FREY, University of California, Davis, USA and Ostrom Workshop, Indiana University,
QIANKUN ZHONG, University of California, Davis, USA
BERIL BULAT, University of California, Davis, USA
WILLIAM D. WEISMAN, University of California, Davis, USA
CAITLYN LIU, University of California, Davis, USA
STEPHEN FUJIMOTO, University of California, Davis, USA
HANNAH M. WANG, University of California, Davis, USA
CHARLES M. SCHWEIK, University of Massachusetts, Amherst, USA

Building a successful community means governing active populations and limited resources. This challenge often requires communities to design formal governance systems from scratch. But the characteristics of successful institutional designs are unclear. Communities that are more mature and established may have more elaborate formal policy systems (as cause or effect of their success). Alternatively, they may require less formalization precisely because of their maturity: because they have more latitude and capacity to select and acculturate new members, or because their reputations encourage greater self-selection. Indeed, scholars often downplay the role that formal rules relative to unwritten rules, norms, and values. But in a community with formal rules, decisions are more consistent, transparent, and legitimate. To understand the relationship of formal institutions to community maturity and governance style, we conduct a large-scale quantitative analysis applying institutional analysis frameworks of self-governance scholar Elinor Ostrom to 80,000 communities across 3 platforms: the sandbox game Minecraft, the MMO game World of Warcraft, and Reddit. We classify communities' written rules according to several institutional taxonomies in order to test predictors of institutional formalization. From this analysis we extract two major findings. First, institutional formalization, the size and complexity of an online community's governance system, is generally positively associated with maturity, as measured by age, population size, or degree of user engagement. Second, we find that online communities employ similar governance styles across platforms, strongly favoring "weak" norms to "strong" requirements. These findings suggest that designers and founders of online communities converge, to some extent independently, on styles of governance practice that are correlated with successful self-governance. With deeper insights into the patterns of successful self-governance, we can help more communities overcome the challenges of self-governance and create for their members powerful experiences of shared meaning and collective empowerment.

# 1 INTRODUCTION

Decades into the emergence of the Internet, online communities continue to thrive and to experiment with new ways of building shared value and meaning. Their successes can be empowering to members, not to mention inspiring, and especially empowering to those users most active in their community's governance. While a community's founding inspiration may be grounded in a personal connection to like-minded peers, or cat memes, maintaining a community requires overcoming daunting governance challenges. A community's governance style and institutional structure therefore are issues of central concern. Governing online communities—wikis, forums, mailing lists, shared servers, BBN's and USENET, MMO's, virtual worlds, open source software projects, and all manner of more modern platforms—is therefore now an important part of several areas of active research in the HCI community, including areas of research focused on peer production [4, 28], online community [36], toxic behavior [37, 42], moderation [21, 38], and open source software [11, 65].

Fortunately, this work is increasingly benefitting from theoretical frameworks in the social sciences, particularly those focused on small, emergent, self-governing communities. The central concern of these frameworks is the challenge of designing institutions to support collective action. As much online as offline, building a successful community requires managing finite, costly resources whose provisioning and distribution raise classic social dilemmas such as the tragedy of the commons [22, 52] and the dilemmas of peer production: attracting volunteer contributions, reducing the cost of free riders, and managing vandalism [39, 50].

Institutions are "the prescriptions that humans use to organize all forms of repetitive and structured interactions" [53]. Institutional Analysis and Development (IAD) is an analytic framework for the empirical analysis of institutions, specifically those focused on natural resource management [53]. It was developed during the 1990s by Elinor Ostrom and the "Ostrom Workshop" community of researchers, which includes economists, resource management scientists, and political scientists and theorists, as well as psychologists, sociologists, and anthropologists. Key ideas in this community are institutional diversity (the idea that successful institutions mix several types of governance regimes) [53], polycentricity (an embrace of policy complexity to match the complexity of the resource environment) [44], and use of research methods that facilitate large-scale comparisons across institutions [5, 52, 58]. By focusing on the finite resources that a community needs to endure and the general threats to managing them sustainably, IAD gains a general handle that has permitted its bold institutional generalizations across pastures, forests, watersheds, fisheries, and now, online communities.

In fact, Ostrom's IAD scholarship has attracted decades of interest from Internet and HCI scholars. Particularly influential have been Ostrom's design principles for the management of common pool resources, a collection of properties shared by enduring resource management institutions that were extracted from a comparison of hundreds of ethnographic studies around the world [9, 52, 77]. An early analysis by Hess decomposed the Internet into several commons [25]. Kollock approached USENET through the resource management lens, explaining emergent solutions to the problem of finite user attention with support from the design principles [35]. Later Viegas et al. analyzed the case of community-led policy formalization on Wikipedia with the design principles as well [75]. In a follow up to this work, Forte et al. [16] extended this analysis to Wikipedia's general policy evolution, coming to the conclusion that although formal policy was adding more kinds of processes and administrators, editors were establishing informal practices that favored decentralized decision making. Even more generally, the wide-ranging work of Kraut and Resnick mapped some of the design principles to their own empirically grounded prescriptions for building successful online communities [36]. More recently, Seering [68] connected the design principles to findings in the content moderation literature, and Frey et al. [19] moved beyond Ostrom's design principles to engage more of IAD's instrumentation for empirical institutional research, using IAD's rule hierarchy to argue for user control over rule-making. Ostrom's thinking has also been fruitfully applied in studies of the World Wide Web [26], USENET [35], online black markets [1], SourceForge [65], and other domains [15, 21].

Institutional analyses of online communities have clearly been productive, but they leave many additional opportunities for online governance insights. Existing work has applied IAD to rich qualitative descriptions of one or a few platforms. Such single-domain inquiries can struggle to capture the full range of potentially effective governance styles (known as institutional diversity [53]). They can also struggle to demonstrate the generality of their findings beyond the small number of communities of interest. In fact, Ostrom explicitly endorsed more "broad but shallow" quantitative applications of IAD as a complement to the more context-sensitive, case-based qualitative research that has dominated institutional research online [58], and emphasized it for large-scale cross-domains comparative research in particular [59]. This has since been echoed explicitly by researchers of online communities [27].

## 1.1 Formal Rule Systems in Online Communities

Policy formalization provides an excellent example of the thrust of contemporary institutional research online. Viegas et al.'s and Forte et al.'s Ostrom-driven investigations of policy evolution on Wikipedia have important implications for all online communities [16, 75], implications that have been taken up in quantitative comparative work on several platforms. In a large study of an ecosystem of fan wikis, Hill and Shaw document the emergence of an oligarchic administrative class [69], while Frey and Sumner show, in a large population of amateur-run servers of the multiplayer game Minecraft, an increase in administrator power with community size and user engagement [19]. One example of attention to written rules specifically is Fiesler et al.'s [14] investigation of written rule systems across Reddit subreddits. They use a topical rule taxonomy ("advertising related,", "bot related", "civility related") to demonstrate context-sensitivity in the variety of rule types that Reddit's diverse communities rely upon.

Whether on Wikipedia or any other platform, communities that are more mature and established, whether in terms of size, age, or depth of user engagement, will have overcome more problems, developed more streamlined practices, and generally settled more deeply into some set of stable sustainable behaviors. We might anticipate that their rules govern a greater range of behaviors in the community, and that a greater proportion of their governance is encoded formally as they become more developed. Certainly this is implied by IAD. Starting with an IAD typology of rules, Basurto and Ostrom [54] define the institutional "state of nature" in terms of default non-rule "rules." Under their account there is, for example, a default access rule that all members can use all resources without constraint, and a default boundary rule that all people can be members of the resource system (Table 1). The idea of institutional defaults is especially interesting as applied to online community platforms, whose software has literal defaults with often unstated governance implications. New Reddit subreddit communities are by default subject to terms of use that broadly define acceptable behavior (conditions of 'access' to discussion "resources," consistent with Ostrom's design principles [9, 77]), but they also, by default, do not set bounds on who may join, on conflict resolution mechanisms, and on other core principles. This implies that communities on this platform cannot expect to attain sustainability without taking the initiative to deviate from software defaults.

Despite the strong claim from IAD-guided research, it is not necessarily obvious that larger, older, or more engaged online communities will have larger or more formal rule systems. In the online realm, the artificial scarcity of many forms of property is particularly artificial (if, with a flip of a bit, a game developer can give players infinite gold), traditionally finite resources like land can be functionally unconstrained (as in the randomly generated, theoretically infinite worlds of Minecraft), and automation is a very real option for assisting in much of the work of an institution's functioning, particularly monitoring and enforcement [7, 12, 13, 29, 78]. With all of these properties in play, it may be possible to support large communities without correspondingly large formal rule systems.

It is important to understand the emergence of formalized policy in online communities for many reasons. One is that governance is costly and can be taxing for moderators and administrators. Scholars have recently paid increasing attention to the hidden costs of moderation [21, 43], as well as to the unintended costs to a community of boundary rules [28]. Distinct from the widely recognized social dilemmas behind common-pool resources and peer production are the "second-order" social dilemmas of maintaining a resource governance system that benefits all and demands support from none [55]. Online examples of the second-order social dilemmas of self-governance include the problems of incentivizing and supporting peer enforcement [32, 43], moderating moderators [20, 57], and even the emotional labor of thanklessly monitoring violations and administering punishments [72]. The cost of behavioral rules is particularly clear for those that regulate the administrator, who must now fulfill the obligations of the institution under a larger set of constraints.

Table 1. "State of nature" default community rules.

| | Default rule |
|---|---|
| 1 | No formal positions exist in the community. |
| 2 | Anyone can enter the community. |
| 3 | Each member can take any physically possible action. |
| 4 | Members act independently. Platform constraints present in the situation determine the aggregation of individual behaviors into outcomes. |
| 5 | Each member can communicate any information via any channel available to the member. |
| 6 | Any member can retain any outcome that the member can obtain and defend. |
| 7 | Each member can affect any state of the community that is physically possible. |

*Source:* Adapted from Ostrom and Basurto (2011)

## 1.2   Contributions

In this work, we investigate the growth of formal rule systems on three platforms for hosting online communities, and the content of the rules that each platform's communities' governments rely on. We look specifically at the growth of formal rules systems as communities become more mature, defined for this work as the increase in community age or population size or depth of user engagement (an operationalization we defend below).

Is there general evidence for a relationship between institutional maturity and formalized governance? Are there cross-platform consistencies in the kinds of governance systems that communities develop for themselves? We perform a quantitative comparative analysis of tens to thousands of communities in three different platforms, 9,800 self-hosted Minecraft servers, 67,000 Reddit communities, and 45 self-hosted pirate World of Warcraft servers—and hundreds of thousands of rules—predicting the size of their formal institutional apparatus, characterizing their governance style, and comparing across domains. Governance in online communities relies on a complex mix of software defaults, custom software, written policies, unspoken norms, and private discussion. This work focuses on only the written component, with some limited attention to governance software. Using the written rules of these communities, and a classifier trained to label them according to prominent institutional taxonomies, we also describe the types of rules that communities tend to rely upon, with a chance to compare governance styles across domains. Our main findings are as follows:

> 1.   *More institutional formalization with more community maturity*: Across platforms, communities that are older, larger, or more engaging have more written rules.

2.      *Cross-platform consistency in governance style*: Online communities use a variety of rule types, and share surprising similarities in their rule type composition across online community platforms, favoring regulative over constitutive policies and norms over strategies and requirements.

With a general comparative approach grounded in interdisciplinary resource management theory, we work towards a research framework for online self-governance in which it is possible to predict when lessons from one type of problem in one kind of community will transfer to different challenges faced by another group, all toward helping online communities across the Internet succeed.

After presenting more background on the three platforms of interest, Minecraft, Reddit, and World of Warcraft, and the IAD's rule taxonomies, we formally introduce our research questions, describe our data, the institutional statement classifiers behind our rule-level analyses, and the statistical models behind our community-level analyses. We close the paper with the implications of this work for our understanding of governing online communities and the opportunities for future research.

## 2   THREE PLATFORMS: PRIVATE MINECRAFT SERVERS, REDDIT SUBREDDIT COMMUNITIES, AND PIRATE WORLD OF WARCRAFT SERVERS

Although the constraints of computer mediated sociality ultimately bind them to a common label, online community platforms differ greatly. We introduce three: Minecraft and its ecosystem of amateur hosted game servers; Reddit and its ecosystem of user moderated "subreddit" (or "sub") forum communities; and World of Warcraft (WoW) and its emergent "private" ecosystem of massively multiplayer role-playing communities that self-host pirate/illegal versions of the game. These three domains—a small-group-focused open virtual world, a community discussion platform, and a massively multiplayer game—capture much of the diversity apparent among online communities, and many of the core similarities. Each consists of large numbers of independent online communities, each defines an administrator class with freedom to form their community's resource governance approach, and researchers of each can automatically extract policies, policy changes, measures of community engagement, and other metadata (Figure 1).

### 2.1   Minecraft

Minecraft is a multiplayer video game that allows players to explore an open world and build creative structures. It is the most prominent of the class of open-ended sandbox and survival games, and one of the most successful video games of all time. Successfully hosting the game involves resolving dilemmas related to in-game resources, as well as common-pool RAM, CPU, bandwidth, and storage resources. To solve these problems, the player community has developed thousands of free software plugins that administrators can use to implement modular governance features, such as whitelists, markets, and administrative hierarchies. Although most research on Minecraft has an educational or media studies character [49], some research has focused on the collective behavior [46, 47] and the self-governance dimension of the game [19, 61, 79]. We leverage a large database of Minecraft server performance and governance data that was used to show that larger and more engaging Minecraft servers have more administrative governance structure [17].

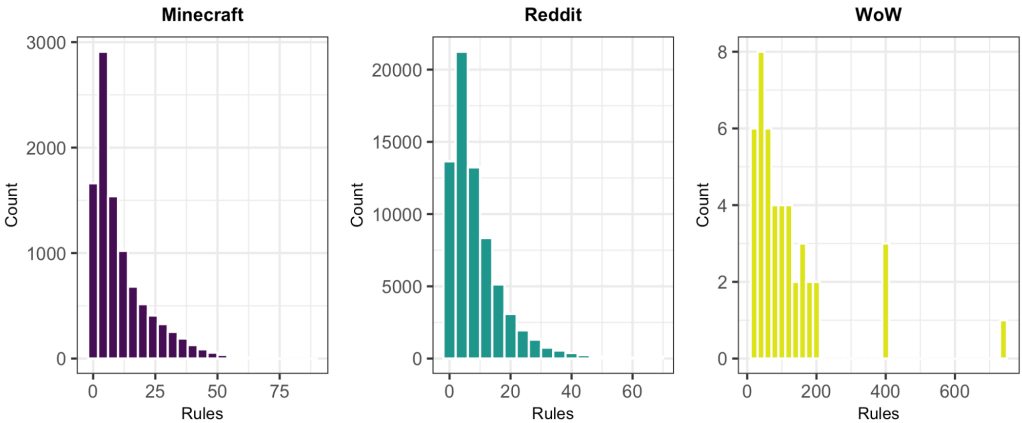

Fig. 1. Communities differ in their level of institutional formalization, within and between platforms. Each histogram shows how many communities have a given number of formal policies, our measure of institutional formalization. Within a platform, most communities are near the median policy count, while some have many more. The different count maxima by platform indicate data constraints, that discussion platform Reddit gave us data on the greatest number of communities, and multiplayer game World of Warcraft (WoW) the least. But the World of Warcraft communities that we did observe had the most policies by far, with the median policy count within WoW close to the maximum count on the other platforms. We operationalize institutional formalization as the number of written rules (WoW and Reddit) or the number of governance-focused software plugins (which can be seen as bundles of rules; Minecraft), which may throw off the comparison. But if it holds, it may be because WoW servers support user populations much larger than those of the game Minecraft, and face resource challenges that are more complex than Reddit's, and unlike Reddit communities cannot benefit from corporate host's pre-existing systemwide terms of service.

## 2.2   Reddit

The discussion site Reddit hosts millions of self-contained "subreddit" discussion communities, for topics from legal advice to pimple-popping. In addition to sitewide policies and standards, each subreddit is governed by volunteer moderators who can write public rules, moderate content, and sanction users [14]. Many researchers interested in amateur governance have taken an interest in the power given to these volunteer moderators [6, 8, 14, 34]. From a resource management perspective, Reddit experiences all of the collective action problems typical of moderation-heavy online discussion platforms: managing bad behavior, encouraging quality content, and the second-order social dilemmas of training and retaining quality moderators [32, 43].

## 2.3   World of Warcraft

World of Warcraft is a massively multiplayer online role playing game (MMORPG), notable for its age (17 years) and enduring popularity. Most institutional research on WoW has focused on the official server ecosystem, with organizational research on the structure of in-game guild communities [33, 48, 76], emergent norms systems [62, 73], and even a natural experiment on epidemic dynamics [40]. But there is a second unofficial private server ecosystem, of perhaps 50 servers at any time, that has attracted a large player base, and all of the governance challenges that come with such success. Research on the pirate ecosystem has focused on its legal aspects [63], as well as cultural dimensions: pirate servers as enclaves for players defying IP restrictions or devoted to community and cooperation [72]. Governance challenges faced by private servers include: managing vandalism; defining, monitoring, and enforcing acceptable behavior; resolving player disputes; and recruiting resources to keep servers live and improving. The institutions that administrators build feature many written policies, and include forum rules, terms of service, refund policies, constraints on player behavior, formal constraints on moderator behavior, and a strong reliance on software support, such as bug trackers repurposed into dispute resolution systems. Compared to Minecraft, WoW's self-hosted ecosystem has fewer servers, many more players per server, and less standardization across servers. Compared to Reddit, WoW has a more complex range of resources to manage, and less support for doing so, since private WoW servers, in contrast to subreddits, do not benefit from the legal and professional resources of a shared professional host. This may help explain why, among other differences, WoW communities have so many more rules on average than communities on the other two platforms.

Table 2. The Regulatory/Constitutive taxonomy of institutional statements.

| Taxonomy | Definition | Example |
|---|---|---|
| Regulatory | Regulates behavior of agents in the institution; often phrased in terms of how an agent can, should, or must behave | *"Players are obligated to report any bugs in the game via the BugTracker.."* *(*WoW: *kronos-wow.com)* |
| Constitutive | Define or introduce or state general aims or values of the institution; can define a term or manifest an entity | *"We are a subreddit dedicated to knowledge about a certain subject with an emphasis on discussion."* *(*Reddit: */r/history)* |

Table 3. The Strategy/Norm/Requirement taxonomy of institutional statements.

| Taxonomy | Definition | Example |
|---|---|---|
| Strategy | Defines or describes possible actions, behaviors, or outcomes. | *"Full rules can be found in the wiki."* *(*Reddit: *r/listentothis)* |
| Norm | Describes what an actor should or must do, or generally what should or must happen. | *"You should: Never share your account information"* *(*WoW: *symmetrywow.com)* |
| Requirement | A Norm with explicitly stated consequences; consequences can be good or bad | *"Any use of Hacked/Moded MC clients will get you a Ban"* *(*Minecraft: *mafianetwork.mcph.co)* |

## 3   GOVERNANCE STYLE, INSTITUTIONAL STATEMENTS AND INSTITUTIONAL TAXONOMIES

Online communities of all types rely on written rules to manage behavior in their communities (Tables 2 and 3). An analysis of rules in Reddit communities by Fiesler and colleagues [14] demonstrates both the widespread use and impressive range and diversity of written rules for community self-governance. We expand this style of inquiry into communities' written policies, while acknowledging the importance of more subtle or interpersonal expressions of a community's institutional structure, such as implicit norms. This goes somewhat counter to a popular focus on the norms and culture of online communities arising from a tendency to idealize the fluid, informal, and social components of community building, and an aversion or disinterest in formal written policies and other regressions to coercive social forms. While that perspective has merit, it is also true that written policies are a major tool of small-scale online institutions, and that, in many communities, they work.

With general institutional taxonomies of rule types, we can compare communities across domains to understand the diversity of governance styles. A major strength of IAD, separate from its rich conceptual apparatus, is the wide range of tools it provides for performing institutional analysis, including frameworks for analyzing rule texts, or, more generally, "institutional statements" (ISs). An institutional statement is a construct of IAD that generalizes the typical understanding of "rule" to include norms and other entirely non-prescriptive articulations of an institution's structure. The full definition of institutional statement is "a shared linguistic constraint or opportunity that prescribes, permits, or advises actions or outcomes for actors (both individual and corporate)" [10, 60, 71]. ISs, therefore, help scholars focus on linguistic expressions of institutional structure, written or unwritten, and construct abstract systems for representing institutions.

Institutional statements can be categorized under several complementary taxonomies. In this work, we highlight two: Regulative/Constitutive and Strategy/Norm/Requirement [51, 53]. The first distinguishes between policies that regulate behavior (by defining and properly incentivizing the range of possible actions) and policies that constitute institutional abstractions (by declaring community values, defining authority positions, and asserting which finite resources require management). The second defines a severity scale, from definitions of possible actions through required actions to actions that are associated with explicit rewards and punishments. We can gain a finer-grained understanding of how online communities are governed by focusing on the kinds of rules communities write.

## 3.1   Regulative/Constitutive

The Regulative/Constitutive (RC) taxonomy is a rough scheme for identifying the target of an IS, which can either put constraints on behavior ("Regulatory") or manifest institutional abstractions ("Constitutive") (Table 2). It derives in part from the speech act theories of philosophers Austin and Searle [2, 66], who expanded the philosophy of language beyond statements with truth value to normative statements and statements that constitute actions in the world ("I do" in a wedding). Regulatory ISs are more familiar in online communities, relating to acceptable behaviors of agents. Constitutive ISs define values or, more often in our context, meaningful vocabulary relevant to behaviors that other ISs regulate. A community with more insider vocabulary and culture, or complex conceptions of acceptable behavior may use constitutive ISs to define the community's terms as part of onboarding. Also, a community intended for a narrow, closely aligned audience may use constitutive statements to define its intended audience and declare its specific values and scope. While the two types of ISs can grow together—imposing severe rules with punishments may require constituting a position in charge of enforcement—examples of their independence are straightforward to imagine. An uncontroversial behavioral regulation that will be widely internalized by many members (say, the expectation that a close-knit group of gamers will converge on a single team chat platform) will not require the constitution of new institutional functions. Conversely, independent of internal-facing processes for behavioral monitoring and enforcement, a community may constitute institutional structures that help it perform outside-facing activities, as in the case of the fund-raising arm of a mission-focused community.

In the context of institutional formalization this distinction permits two different approaches to institutional growth. A regulatory statement's "user-facing" approach highlights an institution's increasingly definite behavioral specification, while a constitutive statement's "self-facing" approach underscores the institution's growth of internal processes, values, positions, and understandings to meet that and other specifications. Naturally, both impose maintenance costs on an institution, but each in different ways. Writing a rule that defines more institutional structure into existence or imposes punishments means devoting more effort (usually volunteer effort) to maintaining all of the institution's process and functions, and ensuring that it operates in the way that it claims to. These include encoding formal policy systems, monitoring and enforcing policy compliance, adjudicating and resolving disputes, recruiting and onboarding users, and also the pedestrian work of educating and informing members about the community's expectations. One hypothesis, not testable in this work, but representative of the potential of the approach, is that online communities will have a higher ratio of regulative to constitutive ISs to the extent that it can automate enforcement (rather than constituting enforcement positions, processes, and functionalities for members to work within). Or it could be that a community will rely more heavily on constitutive statements the more complex it is (managing more kinds of resources under more kinds of users with more kinds of goals). This may describe the communities behind a complex MMORPG like World of Warcraft, in contrast to the more focused and single-topic discussion communities of Reddit.

## 3.2    Strategy/Norm/Requirement

The Strategy/Norm/Requirement (SNR) taxonomy classifies all institutional statements by degree of prescriptiveness or severity, as either Strategy, Norm, or Requirement (Table 3). (The IAD literature speaks of the SNR taxonomy in terms of Rules, Norms, and Strategies. We replace "Rule" with "Requirement" here to avoid further overloading the term). Requirements are the strongest IS in the SNR taxonomy; they are stated with a deontic (e.g. "should" or "must") and attach consequences with an "or else" component, such as a reward or punishment. Norms are one step less restrictive than requirements, containing a deontic but no "or else." So they are consistent with the colloquial sense of "norm" in the sense of being detached from formal enforcement, although they differ because colloquial norms are unwritten and informal, while SNR norms, particularly those we investigate here, are both written and a formal part of the institution. The weakest institutional statement type in this taxonomy is the strategy. Strategies are ISs with no deontic or consequence. They represent the idea that not all "rules" are normative. Strategies can include guidance or "onboarding" style articulations of an actor's choice set. For example, the strategy "*New members can find admin contacts in the FAQ*" does not tell users what they should do, but what they possibly could do.

Community- or domain-level differences in the relative proportions of strategies, norms, and requirements can be straightforwardly understood as differences in governance style, reflecting how much an institution depends on prescriptions and implicit threats in order to ensure compliance. These differences will have implications for how community administrators invest their limited time and energy. Writing a rule that regulates user behavior means putting more effort into monitoring and enforcement (in the case of requirements) or putting effort into instilling internalized compliance (in the case of softer norms). And in the context of institutional formalization, this distinction makes it possible to characterize communities on the severity of their rules, with more closed or culture-driven communities authoring more norms relative to requirements, communities with larger, less personal, or more unruly audiences potentially eliciting more requirements, and communities, large or small, with a greater range of actions and abstractions inviting more strategy statements.

While Reddit likely exhibits this whole range of community types, its private communities may be expected to rely less on requirements than communities that remain public (though, unfortunately, we cannot test this in the present work because the written rules of private communities were not accessible to our scrapers). By default, game communities like Minecraft and World of Warcraft that attract younger or more aggressive or competitive users may be expected to rely more heavily on requirements, rather than internalization.

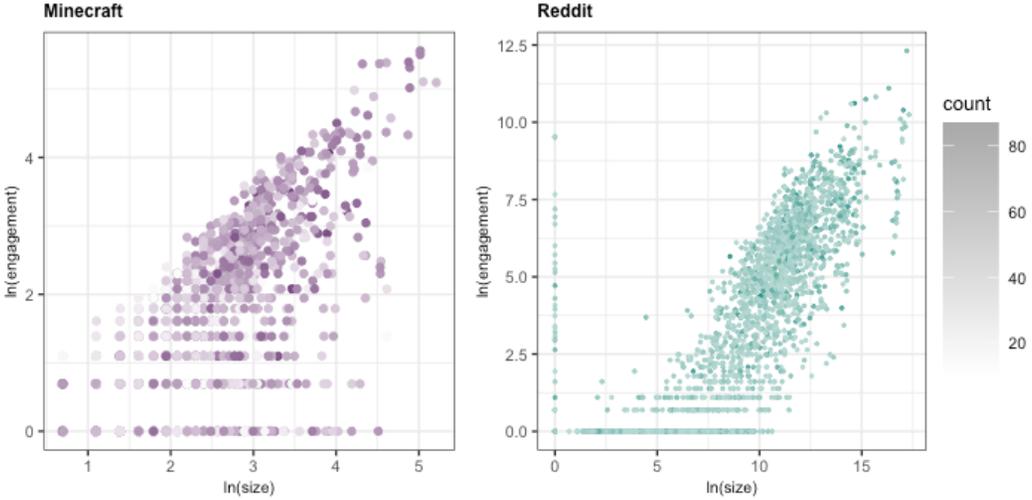

Fig. 2. Size and community engagement are highly correlated, and so suitable as redundant estimators of maturity. Both are associated with greater institutional formalization. This log-log plot shows that the percentage increase in community size is strongly correlated with the percentage increase in user engagement measures from each domain. Engagement and size both increase with policy count (darker points), our measure of formalization. WoW is excluded because its data lack estimates for both size and engagement.

## 4   RESEARCH QUESTIONS

Having established the importance of general insights into online community governance, we next relate measures of **institutional formalization** and **community maturity** and work to understand variation in **governance style**. We measure institutional formalization across domains simplistically in terms of a community's number of formal policies (either written rules or governance plugins in the case of Minecraft; Section 5.1.1), and we represent governance style in terms of the composition of ISs under the RC and SNR taxonomies.

To measure community maturity, we take an opportunistic approach, measuring it in terms of a mix of community age, number of users, *or* depth of user participation, depending on the mix of measures that each platform's communities make available. Our operationalization of maturity, in terms of population size, community age, or depth of user engagement, is motivated by concerns of practicality, operationality, and responsibility. Practically, these properties are empirically correlated (Figure 2) and may therefore be seen as proxies for an umbrella construct like "institutional maturity." Indeed, prior research has used all three to operationalize the sense of advancement relative to other communities [19, 31, 69]. Operationally, the platforms we investigate don't all expose all three dimensions of development, so taking all three together, dependent on the mix of measures that each platform's communities made available, gave us a strategy for making cross-platform comparisons possible. And in terms of responsibility, these three properties are not just highly correlated but mutually causal (age can cause growth can cause engagement can sustain age). Grouping them together as related operationalizations of a single construct, as different faces of a "maturity" coin, helps us sidestep secondary questions of how specifically they relate and cause each other.

Of course, the difficulty around causal attribution extends to the other half of our research question. We are non-committal about the causal direction of any relationship between maturity and formalization: just as maturity may motivate policy creation, strong policy systems support increases in maturity, and we therefore endeavor to avoid causal implications and treat both directions equally.

As for the nature of any correlation, it is most likely that institutions will have greater institutional formalization (more formal policies) the more mature they are. Still, it is possible that formalization impedes or even destabilizes a community through restrictive, onerous, or intimidating levels of structure and bureaucracy, or that computational affordances of the online domain, such as tool-assisted moderation and automated enforcement, reduce the relationship between written policies and maturity. Hence our first research question:

**RQ1**: *What is the relationship between a community's maturity and its level of institutional formalization?*

A distinguishing strength of quantitative approaches is that they are well suited to generalization, permitting us to pose the same question comparatively across the three domains of this study:

**RQ2**: *Does the relationship between maturity and formalization generalize across platforms?*

We also take an interest in the governance styles of our communities and whether those styles differ by platform:

**RQ3**: *What governance styles do communities tend to employ, in terms of types of institutional statements, and how consistent are these styles across domains?*

Following the structure suggested by these questions, our analysis has two main components: a statistical community-level regression analysis interested in correlates of formalization, and a descriptive platform-level comparative analysis according to IAD taxonomies of governance style.

## 5 METHODS

Our analysis examines 1) the relation of maturity and formalization at the community level through statistical models of policy count (our proxy of institutional formalization), domain, and measures of maturity, and 2) institutional statement composition at the domain level, through descriptives that plot breakdowns of IS types by platform. After describing the data from each domain, and our work to make them comparable, we describe our modeling strategy for RQ1 and RQ2, and move onto our process for building rule type classifier for RQ3, through hand-coding of ISs, hand-coding of IS types, model training, validation, and application.

### 5.1 Data

From each domain, we rely upon indirect measures of each community's development and institutional formalization, as well as its written rules. Although we succeeded in collecting data from many communities on each of the three platforms, we emphasize that each made different subsets of relevant data available, such that our measures of community development are not complete for all domains.

*5.1.1 Minecraft.* We based our Minecraft analysis on publicly available longitudinal data on the development and formalization of self-hosted Minecraft servers [17]. This corpus offers several months of observations of policy structure and community development (population size and user engagement, here defined as the number of players who visited a community at least once a week for a month [19]). In Minecraft, communities implement policy using both written rules and software for automated governance. But the balance of governance in Minecraft is implemented via plugins, with written rules playing more of a supporting role. This is partly because of the game's rich plugin API support, and partly because written rules in Minecraft are usually implemented on in-game signage, which is terse (about 60 characters) and used primarily for non-governance game-specific purposes such a object labels ("Hannah's House") and game functions ("For help type /manual"). We therefore base our analysis of institutional formalization on the number of governance-specific software plugins a server has installed. Minecraft governance plugins can be seen as bundles of code-bound rules, a satisfactory operationalization of the institutional formalization construct. We also obtained a sample of written rules from a small subset of these communities, enough, when aggregated within-domain, for our governance style analysis.

*5.1.2 Reddit.* We scraped four types of data from Reddit: generic community metadata from the Reddit API, rule texts from subreddits that use Reddit's "Rules Widget" (67K) as of September 2019, snapshots of subreddit rules archived on the Internet Archive's Wayback Machine (4K), and engagement measures calculated over full September 2019 comment data [3]. Similar to our Minecraft measure, our measure of Reddit community development includes a user engagement measure, this one adapted from bibliometrics. We define engagement in terms of an *i5*: the number of Reddit users who made at least five comments in that subreddit community in a month. While this fails to capture subjective engagement in certain communities (e.g., communities that forbid comments, or r/MechMarket, where users only post or comment when conducting a sale), this measure is consistent with an idea of community grounded in user commitment. From the metadata corpus, we also had access to Reddit community size (number of subscribers) and age.

Restricting our analysis to communities using the Reddit Rules API has some limitations that merit discussion. First, it is a relatively new feature, and optional, and many communities with written rules had not transitioned to the standard by the time of data collection. Before this standard was introduced, communities often concatenated their rulesets in plaintext to the end of their official description on the site (and some persist in this way). This means our data may overrepresent communities that are younger or that started instituting rules later. Additionally, moderators may introduce small differences in how a rule is publicly stated and how it is stated in the context of an enforcement action; where these divergences occurred we made the choice to favor the posted rule-in-form over any contextual variant. Last, unlike in the Minecraft corpus, we did not get access to communities' computationally implemented software policies (e.g., from the Reddit "Automod" feature), whose rules are private by default. Therefore we measured institutional formalization in terms of written rules only, and only a subset of those in effect, some of which may not have been clearly visible to users.

*5.1.3 World of Warcraft.* After combining several lists of private WoW communities (from https://www.dkpminus.com/wow/private-servers/, https://www.reddit.com/r/wowservers/comments/6yjpw3/wow_private_server_list/, and https://www.reddit.com/r/wowservers/comments/37b5ir/wow_private_servers_list/), we manually reviewed each site. Of 124 servers that once existed, 99 left some historical trace, and 70 continue to operate as of September 2020. In all, 66 private servers left rule texts, and 45 of those left sufficient data on their maturity and written rules for regression analysis. Working with both live sites and available archive.org entries, we retrieved each community's age and rule documents. However, these sources did not make measures of community size or engagement available.

## 5.2 Community-level analysis

To investigate RQ1 and RQ2, we regressed institutional formalization against community maturity. Model selection was data-driven, identifying well-fitting models by selecting a subset of variables out of "saturated" models starting with all available covariates: platform, age, population size (log-scaled), engagement and all of their pairwise combinations. We retained covariates on the basis of AIC, breaking ties in favor of models with fewer variables and greater $R^2$ (and remaining blind to *p*-values). By the same approach, we rejected normal linear regression in favor of the negative binomial form, which is consistent with a picture in which communities accumulate policies according to Poisson processes whose specific rates may differ by community. Under this iterative model selection process, we refined our operationalizations, removed non-influential interaction terms, and confirmed the validity of the negative binomial form. For the sake of comparison and redundancy, we converged on five regression models with satisfactory fits: two "general" cross-domain models and three platform-specific models (see Table 4).

Our first cross-domain model regresses on all three community maturity features across both Minecraft and Reddit communities. The second models formalization in all three domains, but only one maturity feature, age (the only covariate available in all three domains). The three domain-specific models play a supplementary role to the main models, providing for robustness and additional interpretation. All of these models have low $R^2$ (around 0.10; Table 4), particularly in the case of the 3-domain model. We tolerate this circumstance on the basis of their low number of parameters and large $n$ [23]. Given the size of our data, the values of $R^2$ we report for each model could not be expected by chance.

During this process we also attempted to address RQ3 statistically, with models of governance style by domain or maturity. These fit particularly poorly, hence our purely descriptive approach to RQ3.

Table 4. Model fit statistics

| Model | Data | df | AIC | θ | $R^2$ |
|---|---|---|---|---|---|
| Model 1 | MC + R | 77306 | 494688 | 1.89 | 0.16 |
| Model 2 | MC + R + WoW | 77356 | 504429 | 1.66 | 0.07 |
| MC-only | MC | 9848 | 67021 | 1.57 | 0.11 |
| Reddit-only | R | 67455 | 427461 | 1.96 | 0.17 |
| WoW-only | WoW | 43 | 519.09 | 1.55 | 0.11 |

## 5.3   Platform-level governance style analysis

To classify our cross-domain corpus of institutional statements according to IAD's Regulatory/Constitutive and Strategy/Norm/Requirement taxonomies, we built and validated a machine classifier for identifying and labeling ISs according to each.

5.3.1   *Training corpus.* Training the classifiers required a corpus of hand-labeled data. The labeling was performed by a team of volunteer data coders who were trained on a codebook and received iterative training sessions. We originally built the hand-labeled corpus with 17 coders from 24,252 rows (2,158 ultimately determined to be ISs), 13,394 from Reddit, 321 from WoW, and 10,537 from Minecraft. After finding poor inter-rater reliability (α=0.08–0.35), we conducted a second, less blind round of analysis by selecting the five coders with pairwise reliability closest to an author-driven master list of labels, for improved values of inter-rater reliability on institutional statements (majority agreement reliability=0.996; Krippendorff's α=0.68), RC (0.995; α=0.65) and SNR (0.99; α=0.51). This resulted in a smaller training corpus of 6422 statements, 4948 of which were ISs. The effects of this second round of coding were higher average agreement between coders, higher agreement with the master labels, and a more accurately trained classifier. We were particularly encouraged that percent agreement ended above 98% for all three codes, indicating that most disagreement was on rare labels. Nevertheless, α remained below Krippendorff's normative standard (α=0.80) and also below the current state of the art for these hand-coded constructs in the field of policy analysis (α=0.735, as calculated from [24]).

Unlike the other domains, which were mostly ISs, only 0.79% (83 rows) of the Minecraft texts were ISs. This was expected, because we extracted text from in-game signage, which is less often used for written rules than other in-game purposes.

5.3.2   *Classifier construction.* We were fortunate to be able to build from the precedent of Rice et al. [60], who offer a pipeline for computational approaches to policy texts, and Fiesler et al. [14], who also trained classifiers over Reddit rules (whose findings we were able to reproduce). Their approach to classifying Reddit rules was based on a system of binary SVMs trained on TF-IDF features of policy texts, including bigrams and trigrams. To accommodate the large number of features, we used a linear SVM implementation designed for scale [56]. We evaluated and validated performance with a 10-fold cross-validation. Because labels in the various taxonomies were imbalanced, we quantified model performance with precision, recall, and F1, all weighted to give less influence to disagreements that occurred on less common labels.

5.3.3 *Analysis pipeline*. After preprocessing (punctuation, whitespace, removing stop words, lowercasing, stemming), the classification tasks were arranged in stages, to first detect institutional statements, and to then classify each detected IS as either regulatory or constitutive, and as either strategy, norm, or requirement.

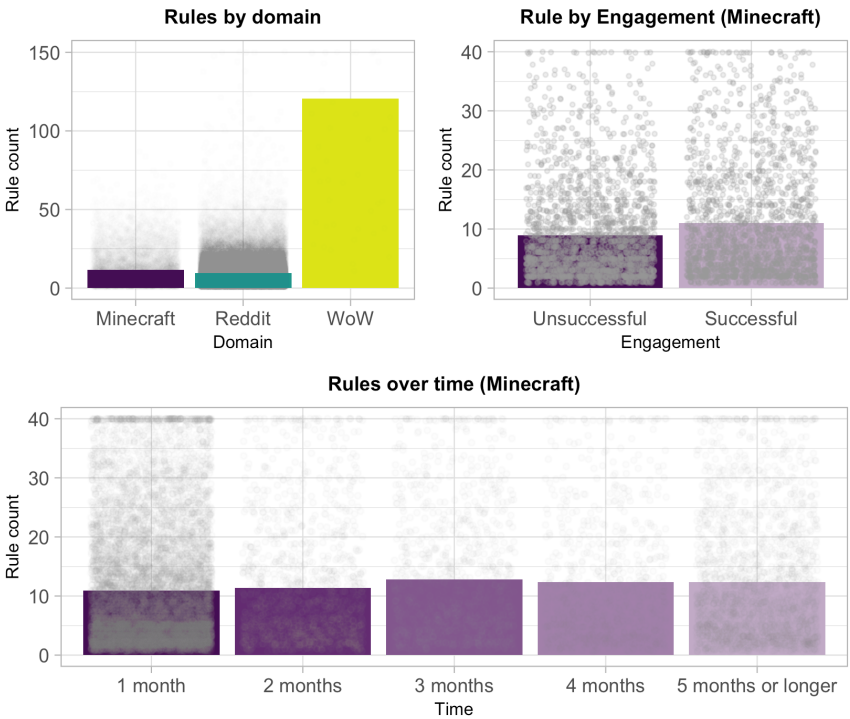

Fig. 3. Communities differ in their level of institutional formalization along dimensions of engagement and time. Among Minecraft servers (right and bottom panels), communities that "succeeded" at user engagement (more than one regularly returning visitor) had more software governance plugins, as did many communities after their first months of existence. The superimposed raw plugin counts (translucent gray) in this figure are truncated at 40 for visualization purposes. The bar hues represent the three platforms; most bars are blue because two of the panels represent Minecraft data only.

## 6 RESULTS

After establishing context with descriptive statistics for each domain, we illuminate RQ1 with a discussion of the regression results, RQ2 with a look to how the models speak to domain-level differences, and RQ3 with an account of our descriptive findings about governance style, in terms of institutional statement type composition by domain.

### 6.1 Descriptives

Our analysis is based on a corpus of 80,296 independent online communities from the three domains. The dataset includes 9,855 servers from Minecraft with a total of 171,796 plugins. From Reddit, 67,462 communities are included, and a total of 667,330 written rules. As for the private WoW, 45 servers with 5,428 written rules are included in our population.

Minecraft servers in our dataset have a median population of 4 (range 1–182), and a median of 7 (1–87) governance plugins. Median depth of user engagement, estimated as the number of community members who return at least once a week within a month, is 1 (0–258; Figure 1). Overall, there are 3,064 servers that "failed" in this sense, with no users visiting on a weekly basis for a month. The median lifetime of a Minecraft server was 2 (1–13) months. Other exploratory descriptive findings support the conclusions of prior work, that older and more engaging communities seem to have more policies (Figure 3).

The 67K Reddit communities display a great range of diversity in community-level features, with a median of 7 written rules (range, 1–87), and median population of 65 (0–2,823,742). Median depth of user engagement in Reddit communities is 0 (0–222,412), which in this case indicates the number of users who posted at least five comments within a month. 52,990 communities "failed" by this standard. The median lifetime of a Reddit community is 30 months (1–176).

For the 45 WoW communities, the median number of institutional statements is 81 (range, 11–732), and the median lifespan of a community is 86 months (1–235). Due to the limitations on available data, we only included variables age and policy count in our analysis.

## 6.2   RQ1: Relating a community's maturity and its level of institutional formalization

In order to gain a comprehensive sense of the interactions of our maturity measures, and differences across domains, we fit several complementary models. The takeaway across these is that community maturity, by many measures, is positively correlated with more formal policy.

**Model 1.** The first model in our analysis considers the correlations of community size, age, engagement, and interactions with the number of policies in each community (Model 1, Table 5). We find significant positive relationships of size, age, and engagement with the numbers of policies communities implement (the main effect of age actually comes out negative but all interactions involving age come out positive; all $p<0.001$, Table 5), suggesting that more established communities tend to have more policies. Model 1 also speaks to RQ2, with the overall finding that maturity and formalization are positively associated in all three domains.

**Model 2.** Our second model supports the Model 1 conclusion for RQ1, that maturity and formalization are positively related. It poses the same hypothesis with fewer maturity features but data from all three domains. In the light of Model 1 it functions to ask whether the outlier platform WoW differs more from the other two domains than they vary from each other (Model 2, Table 5).

**Single-domain models**. We next examine effects from domain-specific models. Overall, the findings from the domain specific models are aligned with those from the combined model: intercepts, effects of size, and most interactions stay significant, are of the same general size, and fit in the same direction as in the combined model.

Table 5: Community size, age, engagement, and time as predictors of policy count

|  | Model 1 | Model 2 | MC-Only | Reddit-Only | WoW-Only |
|---|---|---|---|---|---|
| Intercept | 1.903*** (.017) | 4.195*** (.212) | 1.835*** (.088) | 1.903*** (.011) | 4.195*** (.218) |
| Engagement | .0007*** (<.0001) |  | 0.028*** (.007) | .0007*** (<.0001) |  |
| Age | −.005*** (<.001) | .007** (.002) | −0.027 (.018) | −.005*** (<.001) | .007** (.002) |
| Size | .063*** (.002) |  | 0.279*** (.031) | .064*** (.002) |  |
| Domain (MC) | −.23*** (.028) | −2.248*** (.213) |  |  |  |
| Domain (Reddit) |  | −2.216*** (.212) |  |  |  |
| Domain (MC) x Engagement | .001 (<.001) |  |  |  |  |
| Domain (MC) x Age | .040*** (.005) | .107*** (.006) |  |  |  |
| Domain (MC) x Size | .328*** (.017) |  |  |  |  |
| Domain (Reddit) x Age |  | −.001 (.002) |  |  |  |
| Engagement x Size | > −.001*** (<.001) |  | −.004*** (.001) | > −.001*** (<.001) |  |
| Engagement x Age | <.001** (<.001) |  | −.002*** (<.001) | <.001** (<.001) |  |
| Size x Age | <.001*** (<.001) |  | .035*** (.010) | <.001*** (<.001) |  |
| Observations | 77,317 | 77,362 | 9,855 | 67,462 | 45 |
| Log Likelihood | −494,664 | −504415 | −67004 | −427444 | −513.0 |
| Deviance | 80,333 | 80983 | 10429 | 69934 | 49.3 |

*p<.05, **p<.01, ***p <.001

## 6.3   RQ2: Generalizing community maturity and institutional formalization across domains

Looking by domain, we find that RQ1's general association between maturity and formalization holds across each platform, although they do differ in the strength of the association.

The associations between maturity and formalization suggested by Models 1 and 2 do differ in strength by domain. According to both models, maturity in Minecraft has a stronger relationship with policy count than in the other domains (all $p<0.001$, Table 5), although Reddit and WoW overall show more policies per community than Minecraft (this may be counterbalanced by the fact that a Minecraft plugin is more like a bundle of ISs, suggesting that a "policy" in Minecraft may be worth several policies in the other domains). The major domain-level difference reflected by the domain-specific supporting models is the effect of age, which is positive in WoW, negative in Reddit (both $p<0.001$), and not significant in Minecraft.

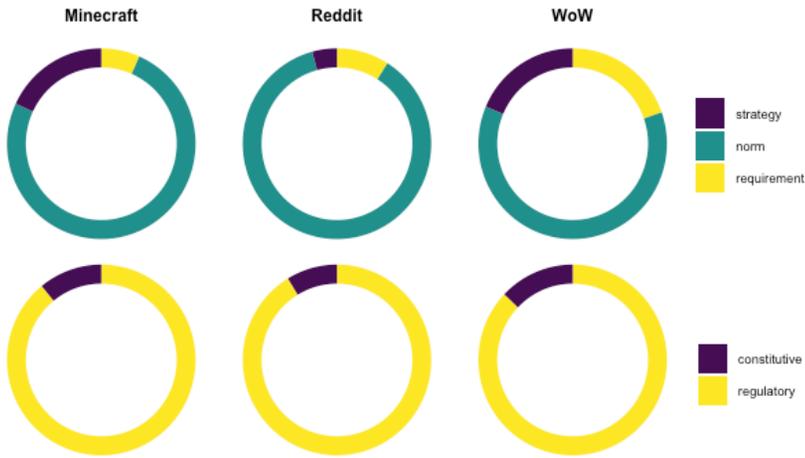

Fig. 4. Under regulative/constitutive and strategy/norm/requirement taxonomies, governance style is consistent across platforms. According to our classification of institutional statements, WoW servers may have a higher dependence on punishment for violations (the difference between a norm and a requirement; first row), and Reddit may use fewer strategy statements than the others, but overall, communities across domains are consistent in the governance regimes they design.

Table 6: Classifier Performance

| Classifier | Accuracy | Precision | Recall | F1 |
|---|---|---|---|---|
| Institutional statement | 0.91 | 0.91 | 0.91 | 0.91 |
| Strategy/Norm/Requirement | 0.88 | 0.86 | 0.88 | 0.87 |
| Regulatory/Constitutive | 0.90 | 0.89 | 0.90 | 0.89 |

### 6.4   RQ3: The governance styles communities employ and their consistency across domains

Domains are surprisingly consistent in the proportions of different institutional statement types that appear in their communities (Figure 4). On all three platforms, regulative ISs strongly outnumber constitutive ISs (accounting for 90.7% of ISs overall), and norms outnumber strategies and requirements by at least two to one. Within the SNR taxonomy 83.5% of ISs are labeled as norms (what users should or must do without explicit consequences), with strategy being the least employed IS type at 8.0%. We noticed a surprising consistency in RC and SNR composition across domains, though this is strictly a qualitative conclusion, as we could not fit a quality model to this data. The success and interpretability of these results depend in large part on the validity of the classifier, which shows strong performance, with accuracy, precision, recall, and F1 scores between 0.86 and 0.91 for all classifiers (Table 6).

If there are significant differences in governance style by domain, they will be that Reddit communities rely less on "strategy" institutional statements, or that WoW communities rely more heavily on requirements. If we see strategies as the non-normative scaffolding on which more normative rules are built, Reddit's seemingly lower proportion of strategies may reflect the fact that it is a simpler domain than either game domain, requiring less explicit articulation of each user's choice set. As for WoW, the greater proportion of requirements suggests a relatively greater role for punishment or sanctions in maintaining compliance than in the other platforms. To this point, WoW game servers are much larger than Minecraft game servers, and they tend to have much more elaborate policy systems and more complex supporting technical infrastructures (forums, appeal systems, moderator hierarchies, and so on).

## 7 DISCUSSION

In the work outlined above, we employ several tools from Elinor Ostrom's Institutional Analysis and Development framework to characterize the governance style of thousands of communities in several platforms, and to demonstrate the co-occurrence of community maturity and formal institutional structure. In doing so, we extend the applicability of institutional analysis to the challenge of building successful online communities. Both the quantity and quality of data make online institutions an ideal testbed for the computational approaches to institutional analysis that we offer.

Our key findings are that online communities that are older, that have larger followings, or that elicit deeper engagement from their members also tend to have larger formal policy systems, not just in terms of computational support like automation and streamlining, but in terms of written rules meant to be read and internalized by new members and used to adjudicate disputes and explain punishments. While communities seem to be able to start with institutional defaults that do not fulfill accepted principles of institutional sustainability, the process of gaining maturity is associated with deviation from these defaults, in the direction of explicit articulations of acceptable user and resource boundaries, user positions, information flows, decision processes, incentives, actions, and outcomes. Indeed, nine out of ten institutional statements are focused more on defining and regulating acceptable behavior than on constituting institutional abstractions such as values, processes, departments, or special positions (Figure 4). They also use norms much more than requirements or strategies, suggesting both a strong reliance on prescribing acceptable behaviors and a general reluctance to back those prescriptions up with punishments.

Of course, an increase in institutional formalization is not necessarily a decrease in the use of informal structure. There is a common understanding of "rules" that contrasts them with soft, non-coercive, context-sensitive "norms," and there is an associated understanding of formal systems as more requirement- than norm- based. None of our findings support this picture. The overwhelming majority of rules that communities write have no explicit coercive component: no clearly associated punishment or consequence. Some written rules are not even normative: they exist merely to articulate to users the actions and opportunities — the strategies — available to them. Because we do not measure informal structure in our dataset along with formal structure, we cannot relate them. Instead, we emphasize that they are as likely complements as substitutes. Forced to choose, we would hypothesize that formal and informal institutional structure are not alternatives, but grow together, and that our findings of an increase in formalization with maturity predict a corresponding increase in complementary "informalization."

But focusing on the effects we actually measure, we are deliberately non-committal about their direction. Does a greater number of explicit formal policies create the framework that communities need to make growth, longevity, and user engagement more possible? Or do growth, longevity, and engagement trigger in communities processes that lead to more formal institutional structure? (If so, is the dominant mechanism going to be bureaucratization, amateurs learning to govern, dictators consolidating power, community adaptations to an evolving environment, or some other institutional change process?)

Either possible direction is interesting and important, and, likely, both are true to at least some extent. But, from the perspective of actionability, the latter is clearly of greater interest, since administrators have more control over their community's policies than its maturity. If an administrator could grow their community merely by writing stronger or more comprehensive policy systems, they would have an easy path to success. The truth is likely more complicated. It may seem that having a larger, more explicit, or more detailed policy system is reassuring because users gain clear steps that they can follow to accomplish their goals in the community. However, research on online communities [36], and Wikipedia in particular [30, 74], has found that online communities with large, complex, comprehensive policy systems are perceived as forbidding, and often require additional welcoming, onboarding, or even mentoring processes in order to smooth out the new user experience enough to retain newcomers. Likely, the appropriate conclusion is that a community and its policies must coevolve, indicating governance design tools that are intended to support iteration, monitoring, and user involvement in higher-level institutional change [18, 64]. By waiting until a new event precipitates an institutional change, administrators can not only ensure that the solution is appropriate to the problem, they can also ensure that the community, having experienced the problem with a lack of policy, will buy into the solution and accept its necessity and, from there, its legitimacy.

## 7.1   Design implications

One reason that less mature communities have less formal structures is that platforms initialize new online communities with "null" governance defaults [54]. Communities that succeed do so only by actively deviating from these defaults. The fact that these deviations share certain properties, not just within but across platforms, suggests that more communities might succeed if they were initialized with more substantive defaults, specifically defaults that follow the governance content and style of existing successful examples. This could look like a range of templates that each community would start with: one set for closed communities dedicated to sensitive topics, another for open communities devoted to lighter topics. An alternative, more flexible, mechanism might permit communities to fork each other's governance.

Of course, one problem with these solutions is that a community and its policies likely coevolve in complexity. Maturity creates the institutional capacity necessary to implement and support new policies, while simultaneously creating the problems that they address. For this reason a new community may be more burdened than benefited by adopting the over-complex policies of a larger peer community. A more nuanced approach to copying policies might be copying trajectories: allowing a new community to adopt the policies that a mature community had when it was new, and to follow those stages as it gains both the means and the needs for greater policy complexity.

Another design implication of these findings is that there is generally much less need in online communities for "hard" requirements with explicit consequences. It is, after all, only a fraction of users who are responsible for most violations [37]. For most users, close monitoring and enforcement for violations is not a replacement for normative guidance, and indeed can interfere with it by undermining affiliative onboarding efforts. Designers may support this need with tools that facilitate onboarding, mentorship, and other "soft" and peer-focused dimensions of community self-governance.

Two-thirds of new admins receive no formal training and "learn based on a combination of understanding the community's values and simply learning by doing" [67], while others receive only domain-specific training and onboarding [43]. With more tools and guidelines for helping regular internet users succeed at the daunting challenges of self-governance, we can support the creation of more kinds of empowering groups online, and provide to more people the power of community.

## 7.2   Limitations

Our multi-domain-spanning, cross-sectional study examined a vast quantity of observational data to derive large-scale, general insights into online institutions and extend the possibilities for offline institutional theories for online community research. However, it is important to note that we compared over only three different domains, and that we retrieved different measures for each domain.

Our analytic framework, the Ostrom Workshop community's Institutional Analysis and Development framework, is flexible and well suited to the online realm. It is designed for comparative analysis, and can bring seemingly incomparable communities together. It also comes with a powerful methodological toolkit. However, IAD has several shortcomings as well. For one, it tends to impose the assumption that a community's successes and failures are due entirely to its governance structure. But a community's policy systems are undoubtedly influenced by numerous factors outside the scope of resource management problems. A community's topic or broader cultural norms and standards will certainly influence its approach to governance. While a free-wheeling or controversial community may commit to minimal content standards, a community covering legally sensitive subjects—drugs, abuse, self-harm—may lean more heavily on explicit ground rules, and the idea of virality seems to allow success to precede governance. Other critiques of IAD target its relationship to rational choice, its methodological individualism, its interest in mechanism and structure over deliberation and personal transformation, and its incomplete integration of critical perspectives on social power [41, 45, 70].

The greatest room for improvement is in our analysis of governance style. General schemes for classifying the types of policies that communities implement will be invaluable for providing communities with actionable governance best practices, and for better understanding institutional structure as a predictor of success. But more intersubjectively consistent rule type definitions are important for improving the reliability of these constructs, and better theory for interpreting rule type results will be essential for them to be able to contribute meaningfully to governance issues as they are understood in the HCI community. Relatedly, our use of written policies excludes unspoken norms, and our analysis was conducted without any sense of the overlap between the "rules in form" on which we base our analysis and the "rules in use" on which each community actually functions, nor on unobserved governance settings like private settings and messages.

## 8   CONCLUSION

The Internet is a vast laboratory of experiments in governance, self-governance, and self-government's interactions with technology. Although most attention to governance online is commanded by technology policy scholarship and social media giants, there is a sea of small actors trying to build a community around personal interests and discovering the importance and value of formal policies and structure. By better understanding how amateur administrators govern and how effectively their successes transfer, we can help support the ecosystem of online communities and increase the dynamism and empowering potential of the Internet.

## ACKNOWLEDGEMENTS

The authors wish to thank members of the institutionalgrammar.org, communitydata.science, and metagov.org research communities; Kabir Sahni and Jingyuan Liang for help on the WoW data; Simon Marti for the Minecraft sign scraper; and research assistants Jingying Jiang, Jiemin Huang, Baani Dhanoa, Edward Chew, Sabba Beizai, Merihan Daniel, Madelyn Hart, Jennifer Pascalis, Franklin Rodas, and Yongying Ye. This work was supported by NSF RCN 1917908: "Coordinating and Advancing Analytical Approaches for Policy Design" and NSF GCR 2020751 "Jumpstarting Successful Open-Source Software Projects With Evidence-Based Rules and Structures."

## REFERENCES

[1]   Afroz, S. et al. 2013. Honor Among Thieves: A Commons Analysis of Cybercrime Economies. *2013 APWG eCrime Researchers Summit* (Sep. 2013), 1–11.
[2]   Austin, J.L. 1975. *How to Do Things with Words*. Clarendon Press, Oxford, United Kingdom.
[3]   Baumgartner, J. et al. 2020. The Pushshift Reddit Dataset. *Proceedings of the International AAAI Conference on Web and Social Media* (2020), 830–839.
[4]   Benkler, Y. et al. 2015. Peer Production: A Form of Collective Intelligence. *Handbook of Collective Intelligence*. M.S. Bernstein and T.W. Malone, eds. MIT Press. 175–204.
[5]   Cardenas, J.-C. et al. 2013. Dynamics of Rules and Resources: Three New Field Experiments on Water, Forests and Fisheries. *Handbook on Experimental Economics and the Environment*. John A. List and Michael K. Price, ed. Edward Elgar Publishing, Cheltenham, United Kingdom. 319–345.
[6]   Centivany, A. and Glushko, B. 2016. "Popcorn Tastes Good" Participatory Policymaking and Reddit's. *Proceedings of the 2016 CHI Conference on Human Factors in Computing Systems* (2016), 1126–1137.
[7]   Chandrasekharan, E. et al. 2019. Crossmod: A Cross-Community Learning-based System to Assist Reddit Moderators. *Proceedings of the ACM on Human-Computer Interaction* (New York, NY, USA, Nov. 2019), 1–30.

[8]    Chandrasekharan, E. et al. 2018. The Internet's Hidden Rules: An Empirical Study of Reddit Norm Violations at Micro, Meso, and Macro Scales. *Proceedings of the ACM on Human Computer Interaction* (New York, NY, USA, Nov. 2018), 1–25.

[9]    Cox, M. et al. 2010. A Review of Design Principles for Community-based Natural Resource Management. *Ecology and Society*. 15, 4 (2010), 37–56.

[10]   Crawford, S.E.S. and Ostrom, E. 1995. A Grammar of Institutions. *The American political science review*. 89, 3 (1995), 582–600. DOI:https://doi.org/10.2307/2082975.

[11]   Crowston, K. and Howison, J. 2005. The Social Structure of Free and Open Source Software Development. *First Monday*. 10 (Feb. 2005), 2. DOI:https://doi.org/10.5210/fm.v10i2.1207.

[12]   De Filippi, P. et al. 2021. Modular Politics: Toward a Governance Layer for Online Communities. *Proceedings of the ACM on Human-Computer Interaction*. 5, (Apr. 2021), 1–26. DOI:https://doi.org/10.1145/3449090.

[13]   Fan, J. and Zhang, A.X. 2020. Digital Juries: A Civics-Oriented Approach to Platform Governance. *Proceedings of the 2020 CHI Conference on Human Factors in Computing Systems* (New York, NY, USA, Apr. 2020), 1–14.

[14]   Fiesler, C. et al. 2018. Reddit rules! Characterizing an ecosystem of governance. *Twelfth International AAAI Conference on Web and Social Media* (2018), 72–81.

[15]   Fish, A. et al. 2011. Birds of the Internet. *Journal of Cultural Economy*. 4, 2 (May 2011), 157–187. DOI:https://doi.org/10.1080/17530350.2011.563069.

[16]   Forte, A. et al. 2009. Decentralization in Wikipedia Governance. *Journal of Management Information Systems*. 26, 1 (Jul. 2009), 49–72. DOI:https://doi.org/10.2753/MIS0742-1222260103.

[17]   Frey, S. 2019. Emergence of Complex Institutions in a Large Population of Self-Governing Communities. *Emergence of integrated institutions in a large population of self-governing communities*. PLoS One.

[18]   Frey, S. et al. 2019. "This Place Does What It Was Built For": Designing Digital Institutions for Participatory Change. *Proceedings of the ACM on Human-Computer Interaction* (New York, NY, USA, Nov. 2019), 1–31.

[19]   Frey, S. and Sumner, R.W. 2019. Emergence of Integrated Institutions in a Large Population of Self-governing Communities. *PloS one*. 14, 7 (Jul. 2019), e0216335. DOI:https://doi.org/10.1371/journal.pone.0216335.

[20]   Ghosh, A. et al. 2011. Who Moderates the Moderators? Crowdsourcing Abuse Detection in User-generated Content. *Proceedings of the 12th ACM Conference on Electronic Commerce* (New York, NY, United States, 2011).

[21]   Grimmelmann, J. 2015. The Virtues of Moderation. *Yale JL & Tech.* (2015), 42–109. DOI:https://doi.org/10.31228/osf.io/qwxf5.

[22]   Hardin, G. 1968. The Tragedy of the Commons. The Population Problem Has No Technical Solution; It Requires a Fundamental Extension in Morality. *Science*. 162, 3859 (Dec. 1968), 1243–1248.

[23]   Harrell, F.E., Jr 2015. *Regression Modeling Strategies: With Applications to Linear Models, Logistic and Ordinal Regression, and Survival Analysis*. Springer.

[24]   Heikkila, T. and Weible, C.M. 2018. A semiautomated approach to analyzing polycentricity. *Environmental policy and governance*. 28, 4 (Jul. 2018), 308–318. DOI:https://doi.org/10.1002/eet.1817.

[25]   Hess, C. 1995. The virtual CPR: the internet as a local and global common pool resource. *Reinventing the Commons," the fifth annual conference of the International Association for the Study of Common Property, Bodoe, Norway* (1995).

[26]   Hess, C. 1996. Untangling the Web: The Internet as a Commons. *Workshop in Political Theory and Policy Analysis, Indiana University* (1996).

[27]   Hill, B.M. and Shaw, A. 2020. Studying Populations of Online Communities. *The Oxford Handbook of Networked Communication*. B.F.W.A. González-Bailón, ed. Oxford University Press.

[28]   Hill, B.M. and Shaw, A. 2020. The Hidden Costs of Requiring Accounts: Quasi-Experimental Evidence From Peer Production. *Communication Research*. 48, 6 (May 2020), 771–795. DOI:https://doi.org/10.1177/0093650220910345.

[29]   Jhaver, S. et al. 2019. Human-Machine Collaboration for Content Regulation: The Case of Reddit Automoderator. *ACM Trans. Comput.-Hum. Interact*. 26, 5 (Jul. 2019), 1–35. DOI:https://doi.org/10.1145/3338243.

[30]   Jullien, N. et al. 2015. The rise and fall of an online project: is bureaucracy killing efficiency in open knowledge production? *Proceedings of the 11th International Symposium on Open Collaboration* (New York, NY, USA, Aug. 2015), 1–10.

[31]   Kairam, S.R. et al. 2012. The life and death of online groups: predicting group growth and longevity. *Proceedings of the 5th ACM International Conference on Web Search and Data Mining* (New York, NY, USA, Feb. 2012), 673–682.

[32]   Kerr, A. and Kelleher, J.D. 2015. The Recruitment of Passion and Community in the Service of Capital: Community Managers in the Digital Games Industry. *Critical Studies in Media Communication*. 32, 3 (2015), 177–192. DOI:https://doi.org/10.1080/15295036.2015.1045005.

[33]   Kiene, C. et al. 2018. Managing Organizational Culture in Online Group Mergers. *Proceedings of the ACM on Human-Computer Interaction* (2018), 1–21.

[34]   Kiene, C. et al. 2019. Technological Frames and User Innovation: Exploring Technological Change in Community Moderation Teams. *Proc. ACM Hum.-Comput. Interact*. 3, CSCW (Nov. 2019), 1–23. DOI:https://doi.org/10.1145/3359146.

[35]   Kollock, P. and Smith, M. 1996. Managing the Virtual Commons. *Computer-mediated Communication: Linguistic, Social, and Cross-cultural Perspectives*. S.C. Herring, ed. John Benjamins Publishing Company. 109–128.

[36]   Kraut, R.E. and Resnick, P. 2012. *Building Successful Online Communities: Evidence-Based Social Design*. MIT Press.

[37]   Kumar, S. et al. 2017. Antisocial Behavior on the Web: Characterization and Detection. *Proceedings of the 26th International Conference on World Wide Web Companion* (Republic and Canton of Geneva, CHE, Apr. 2017), 947–950.

[38]   Lampe, C. et al. 2014. Crowdsourcing Civility: A Natural Experiment Examining the Effects of Distributed Moderation in Online Forums. *Government Information Quarterly*. 31, 2 (Apr. 2014), 317–326. DOI:https://doi.org/10.1016/j.giq.2013.11.005.

[39]   Ledyard, J.O. 1994. Public Goods: A Survey of Experimental Research. *The Handbook of Experimental Economics*. 861 (Feb. 1994), 111–194.

[40]   Lofgren, E.T. and Fefferman, N.H. 2007. The Untapped Potential of Virtual Game Worlds to Shed Light on Real World Epidemics. *The Lancet Infectious Diseases*. 7, 9 (2007), 625–629. DOI:https://doi.org/10.1016/s1473-3099(07)70212-8.

[41]   Lowery, D. et al. 1992. Citizenship in the Empowered Locality. *Urban affairs quarterly*. 28, 1 (1992), 69–103. DOI:https://doi.org/10.1177/004208169202800104.

[42]   Marlin-Bennett, R. and Thornton, E.N. 2012. Governance Within Social Media Websites: Ruling New Frontiers. *Telecommunications Policy*. 36, 6 (Jul. 2012), 493–501. DOI:https://doi.org/10.1016/j.telpol.2012.01.002.

[43]   Matias, J.N. 2019. The Civic Labor of Volunteer Moderators Online. *Social Media Society*. 5, 2 (2019). DOI:https://doi.org/10.1177/2056305119836778.

[44]   McGinnis, M.D. and Indiana University, Bloomington. Workshop in Political Theory and Policy Analysis 1999. *Polycentricity and Local Public Economies: Readings from the Workshop in Political Theory and Policy Analysis*. University of Michigan Press.

[45]   Mollinga, P.P. 2001. Water and Politics: Levels, Rational Choice and South Indian Canal Irrigation. *Futures*. 33, 8 (Oct. 2001), 733–752. DOI:https://doi.org/10.1016/S0016-3287(01)00016-7.

[46]   Müller, S. et al. 2015. HeapCraft Social Tools: Understanding and Improving Player Collaboration in Minecraft. *Proceedings of the 10th International Conference on the Foundations of Digital Games (FDG)* (2015).

[47]   Müller, S. et al. 2015. Statistical Analysis of Player Behavior in Minecraft. *Proceedings of the 10th International Conference on the Foundations of Digital Games* (Santa Cruz, CA, USA, Dec. 2015).

[48]   Nardi, B. and Harris, J. 2006. Strangers and Friends. *Proceedings of the 2006 20th Anniversary Conference on Computer Supported Cooperative Work* (New York, NY, United States, 2006).

[49]   Nebel, S. et al. 2016. Mining learning and crafting scientific experiments: a literature review on the use of minecraft in education and research. *British journal of educational technology: journal of the Council for Educational Technology*. 19, 2 (2016), 355–366.

[50]   Olson, M. 1965. *The Logic of Collective Action: Public Goods and the Theory of Groups*. Harvard University Press.

[51]   Ostrom, E. 2010. Beyond markets and states: Polycentric governance of complex economic systems. *The American economic review*. 100, 3 (Jun. 2010), 641–672. DOI:https://doi.org/10.1257/aer.100.3.641.

[52]   Ostrom, E. 1990. *Governing the Commons: The Evolution of Institutions for Collective Action*. Cambridge University Press.

[53]   Ostrom, E. 2006. *Understanding Institutional Diversity*. Princeton University Press.

[54]   Ostrom, E. and Basurto, X. 2011. Crafting analytical tools to study institutional change. *Journal of Institutional Economics*. 7, 3 (Sep. 2011), 317–343. DOI:https://doi.org/10.1017/S1744137410000305.

[55]   Panchanathan, K. and Boyd, R. 2004. Indirect reciprocity can stabilize cooperation without the second-order free rider problem. *Nature*. 432, 7016 (Nov. 2004), 499–502. DOI:https://doi.org/10.1038/nature02978.

[56]   Pedregosa, F. et al. 2011. Scikit-learn: Machine learning in Python. *The Journal of Machine Learning Research*. 12, (2011), 2825–2830.

[57]   Poor, N. 2005. Mechanisms of an Online Public Sphere: The Website Slashdot. *Journal of Computer-Mediated Communication*. 10, 2 (2005). DOI:https://doi.org/10.1111/j.1083-6101.2005.tb00241.x.

[58]   Poteete, A.R. et al. 2010. *Working Together: Collective Action, the Commons, and Multiple Methods in Practice*. Princeton University Press.

[59]   Poteete, A.R. and Ostrom, E. 2008. Fifteen Years of Empirical Research on Collective Action in Natural Resource Management: Struggling to Build Large-N Databases Based on Qualitative Research. *World development*. 36, 1 (Jan. 2008), 176–195. DOI:https://doi.org/10.1016/j.worlddev.2007.02.012.

[60]   Rice, D. et al. 2021. Machine Coding of Policy Texts with the Institutional Grammar. *Public administration*. 99, 2 (Jan. 2021), 248–262. DOI:https://doi.org/10.1111/padm.12711.

[61]   Rolfes, L. and Passig, K. 2019. The Proto-Governance of Minecraft Servers. *Journal for Virtual Worlds Research*. 12, 3 (2019). DOI:https://doi.org/10.4101/jvwr.v12i3.7365.

[62]   Ross, T.L. and Collister, L.B. 2014. A Social Scientific Framework for Social Systems in Online Video Games: Building a Better Looking for Raid Loot System in World of Warcraft. *Computers in Human Behavior*. 36, (Jul. 2014), 1–12. DOI:https://doi.org/10.1016/j.chb.2014.03.023.

[63]   Schlüter, M. 2017. *Commoning Games? The Case of the World of Warcraft Private Server Nostalrius*. Leuphana University.

[64]   Schneider, N. et al. 2021. Modular Politics: Toward a Governance Layer for Online Communities. *Proceedings of the ACM on Human-Computer Interaction* (New York, NY, USA, Apr. 2021), 1–26.

[65]   Schweik, C.M. and English, R.C. 2012. *Internet Success: A Study of Open-source Software Commons*. MIT Press.

[66]   Searle, J.R. 1995. *The Construction of Social Reality*. Simon and Schuster.

[67]    Seering, J. et al. 2019. Moderator Engagement and Community Development in the Age of Algorithms. *New Media & Society*. 21, 7 (2019), 1417–1443. DOI:https://doi.org/10.1177/1461444818821316.

[68]    Seering, J. 2020. Reconsidering Self-Moderation: the Role of Research in Supporting Community-Based Models for Online Content Moderation. *Proc. ACM Hum.-Comput. Interact.* 4, CSCW2 (Oct. 2020), 1–28. DOI:https://doi.org/10.1145/3415178.

[69]    Shaw, A. and Hill, B.M. 2014. Laboratories of Oligarchy? How the Iron Law Extends to Peer Production. *The Journal of communication*. 64, 2 (Mar. 2014), 215–238. DOI:https://doi.org/10.1111/jcom.12082.

[70]    Shrestha, K.K. and Ojha, H.R. 2017. Chapter 2 - Theoretical Advances in Community-Based Natural Resources Management: Ostrom and Beyond. *Redefining Diversity & Dynamics of Natural Resources Management in Asia, Volume 1*. G.P. Shivakoti et al., eds. Elsevier. 13–40.

[71]    Siddiki, S. et al. 2019. Institutional Analysis with the Institutional Grammar. *Policy studies journal: the journal of the Policy Studies Organization*. 18, (Jul. 2019), 643. DOI:https://doi.org/10.1111/psj.12361.

[72]    Steiger, M. et al. 2021. The Psychological Well-Being of Content Moderators: The Emotional Labor of Commercial Moderation and Avenues for Improving Support. *Proceedings of the 2021 CHI Conference on Human Factors in Computing Systems* (New York, NY, USA, May 2021), 1–14.

[73]    Strimling, P. and Frey, S. 2020. Emergent Cultural Differences in Online Communities' Norms of Fairness. *Games and Culture*. 15, 4 (2020), 394–410. DOI:https://doi.org/10.1177/1555412018800650.

[74]    Suh, B. et al. 2009. The singularity is not near: slowing growth of Wikipedia. *Proceedings of the 5th International Symposium on Wikis and Open Collaboration* (New York, NY, USA, Oct. 2009), 1–10.

[75]    Viégas, F.B. et al. 2007. The Hidden Order of Wikipedia. *Online Communities and Social Computing* (2007), 445–454.

[76]    Williams, D. et al. 2006. From Tree House to Barracks. *Games and Culture*. 1, 4 (2006), 338–361. DOI:https://doi.org/10.1177/1555412006292616.

[77]    Wilson, D.S. et al. 2013. Generalizing the Core Design Principles for the Efficacy of Groups. *Journal of economic behavior & organization*. 90, (Jun. 2013), S21–S32. DOI:https://doi.org/10.1016/j.jebo.2012.12.010.

[78]    Zhang, A.X. et al. 2020. PolicyKit: Building Governance in Online Communities. *UIST '20: Proceedings of the 33rd Annual ACM Symposium on User Interface Software and Technology* (Aug. 2020), 365–378.

[79]    Zhong, Q. and Frey, S. 2020. Institutional Similarity Drives Cultural Similarity among Online Communities. *arXiv [cs.SI]*.